
\def\pd{\partial}
\def\ts{\thinspace}
\magnification=1200
\pageno=0
\parskip 3 pt plus 1pt minus 1 pt
\rightline{DTP-92/37}
\rightline{July 1992}
\vskip 2 true cm
\centerline{A QUANTUM ANOMALY FOR RIGID PARTICLES}
\vskip 2.5 true cm
\centerline{JAN GOVAERTS%
\footnote{$\sp {\dag}$}
{\rm{Address {}from 1st October 1992:}\hfill\break
\it{Institut de Physique Nucl\'eaire, Universit\'e Catholique de
Louvain,\hfill\break
B-1348 Louvain-la-Neuve (Belgium)}}}
\vskip 0.5 true cm
\centerline{\it{Department of Mathematical Sciences}}
\centerline{\it{University of Durham, Durham DH1 3LE, UK}}
\vskip 2 true cm
\centerline{Abstract}
\vskip 1 true cm
Canonical quantisation of rigid particles is considered paying special
attention to the restriction on phase space due to causal propagation.
A mixed Lorentz-gravitational anomaly is found in the commutator of Lorentz
boosts with world-line reparametrisations. The subspace of gauge invariant
physical states is therefore not invariant under Lorentz transformations.
The analysis applies for an arbitrary extrinsic curvature dependence
with the exception of only one case to be studied separately.
Consequences for rigid strings are also discussed.
\vfill\eject
\vskip 20pt
\leftline{\bf 1. Introduction}
\vskip 15pt
Following Polyakov's proposal$\sp {[1,2]}$ of adding to the ordinary
Nambu-Goto action of string theory a scale invariant term quadratic in the
world-sheet extrinsic curvature -- as a means of smoothening out the
world-sheet structure at short distances, and possibly also obtaining a
low-energy effective theory for QCD --, there has been a revival of interest
in theories of relativistic particles whose action includes a dependence
on the world-line extrinsic curvature and torsion, so called rigid
particles$\sp {[3-18]}$. These theories not only provide a simpler setting
in which to study some of the difficulties presented by rigid strings
due to their nonlinear higher derivative actions, especially at the
quantum level, but it has also been suggested that rigid particles could
define a unifying formalism for quantum particles of arbitrary spin using
spacetime coordinates only (even half-integer spins have been
claimed$\sp {[6,10]}$, though erroneously$\sp {[6,19]}$, to be possible).
However, the understanding of quantum rigid particles presently available
in the literature is not really satisfactory, being rather confusing and
even self-contradictory at times.

In view of this situation, the present letter reports on results of a
detailed analysis of the canonical quantisation of rigid particles whose
action includes an {\it arbitrary\ts} dependence on the world-line extrinsic
curvature but is independent of its extrinsic torsion. Leaving details of
the analysis to a separate publication$\sp {[20]}$, the main emphasis here
is on the existence of a Lorentz anomaly for physical states. Namely, at the
quantum level, while both the local world-line gauge algebra and the
spacetime Poincar\'e algebra are each anomaly free, world-line
reparametrisations {\it do not} commute with Lorentz boosts.
Consequently, quantum physical states {\it do not} transform covariantly
under the Lorentz group. Even though the system only has a finite number of
degrees of freedom, the associated anomaly arises due to operator
ordering and the need to perform a certain map from the original
restricted phase space to an unrestricted one. The initial restriction
originates from the requirement that classical world-line trajectories are
time-like and causal, namely that the velocity of the particle is always less
than the speed of light so that its trajectories always lie {\it inside} the
light-cone (in spite of this, classical solutions always
include$\sp {[6,12,13,20]}$ {\it tachyonic} but nevertheless causal ones).
The present anomaly is quite analoguous to the mixed triangular anomaly
in four dimensions for two gravitons and one U(1) gauge boson$\sp {[21]}$,
Poincar\'e invariance playing the r$\hat{\rm o}$le of an internal symmetry
for rigid particles.

The discussion outlined in this letter applies whatever the dependence on
the extrinsic curvature, with one exception however referred to as
``the degenerate case''. In contradistinction with the
``generic case'', the degenerate case possesses$\sp {[6,20]}$ additional
first- and second-class constraints$\sp {[22]}$ and actually requires an
independent analysis using more advanced techniques. Having not been
completed yet, the discussion of the degenerate case is not included here.
Nevertheless, no fundamental difference with the generic case is to be
expected. The same type of anomaly as the one described above for the
generic case would presumably appear in the quantisation of degenerate
rigid particles.
\vfill\eject
\vskip 20pt
\leftline{\bf 2.  Classical Rigid Particles}
\vskip 15pt
Consider the following action for rigid particles
$$S[x\sp \mu,q\sp \mu,\lambda\sp \mu]=\ -\mu c \int\sp {\tau_f}_{\tau_i}
d\tau\bigl[\sqrt{-q\sp 2}\ F(\kappa\sp 2 K\sp 2)\ +
\ \lambda_\mu(q\sp \mu-\dot x\sp \mu)\bigr]\ .\eqno(2.1)$$
These particles, whose trajectories are described by functions
$x\sp \mu(\tau)$ of the world-line parameter $\tau$, propagate in a
$D$-dimensional Minkowski spacetime with metric $\eta_{\mu\nu}$ of signature
mostly plus signs (as usual, dots above quantities refer to $\tau$
derivatives and $(\mu,\nu=0,1,\cdots,D-1)$\ts). The system is characterized
by two fundamental positive constants $\mu$ and $\kappa$ with dimensions of
mass and length respectively, while $c$ denotes the speed of light. The
coordinates $x\sp \mu$ have dimension of length and $\tau$ is taken to be
dimensionless. The overall sign in (2.1) is a matter of convention$\sp {[22]}$,
chosen such that for $F(\kappa\sp 2 K\sp 2)$ constant and positive, solutions
of positive energy propagate forward in time thus describing particles as
opposed to antiparticles. The dimensionless variables $\lambda\sp \mu(\tau)$
are Lagrangian  multipliers for the constraints
$q\sp \mu(\tau)=\dot x\sp \mu(\tau)$ identifying the degrees of freedom
$q\sp \mu(\tau)$ with the particle velocity. Finally,
$F(\kappa\sp 2 K\sp 2)$ is a given dimensionless non constant
function specifying the dependence of the action on the world-line extrinsic
curvature
$$K\sp \mu=\ {(q\dot q)q\sp \mu-q\sp 2\dot q\sp \mu\over (q\sp 2)\sp 2}\ ,
\quad
K\sp 2=\ {q\sp 2\dot q\sp 2-(q\dot q)\sp 2\over(q\sp 2)\sp 3}\ \ .\eqno(2.2)$$
When using $q\sp \mu=\dot x\sp \mu$, we have indeed the correspondence
$$K\sp \mu=\ {d n\sp \mu\over d s}=
\gamma\sp {-1/2}{d\over d \tau}\bigl[\gamma\sp {-1/2} \dot x\sp \mu\bigr]
\ ,\quad
n\sp \mu={d x\sp \mu\over d s}=\gamma\sp {-1/2}\dot x\sp \mu\ ,\eqno(2.3)$$
where $n\sp \mu$ is the normalised world-line tangent vector,
$\gamma=-\dot x\sp 2$ the induced world-line metric and $d s$ the
proper-time line element $d s=\gamma\sp {1/2} d\tau$.

In defining the action, it is understood that only {\it time-like}
trajectories are to be considered, corresponding to the restriction
$(-q\sp 2 > 0)$ and implying a causal propagation inside the local light-cone
(the unphysical case of supraluminal rigid particles has also been
considered in the literature$\sp {[6]}$). Consequently, we have
$$n\sp 2=-1\ ,\quad K\sp 2 > 0\ ,\quad n K = 0\ ,\eqno(2.4)$$
showing that any function $F(x)$ which is well defined for positive
arguments is acceptable in the definition of (2.1). Nevertheless,
classical {\it tachyonic} solutions of constant extrinsic curvature
always exist$\sp {[6,12,13,20]}$ for arbitrary choices of
$F(\kappa\sp 2 K\sp 2)$.

The degenerate case mentioned in the introduction corresponds to the function
$$F(x)=\ \alpha_0\sqrt{x}+\beta_0\ .\eqno(2.5)$$
However, solutions then exist only$\sp {[6,12,20]}$ when $\beta_0\neq 0$ and
$\alpha_0/\beta_0 > 0$, in which case all solutions are$\sp {[12]}$ of
constant extrinsic curvature and include tachyonic ones for curvatures
$\sqrt{\kappa\sp 2 K\sp 2} > \beta_0/\alpha_0$. Any other choice for
$F(\kappa\sp 2 K\sp 2)$ different from (2.5) defines the generic
case to which the present discussion applies. Only degenerate rigid
particles require a separate treatment not included here.

By construction, (2.1) is invariant under world-line reparametrisations
-- including orientation reversing ones -- with the coordinates $x\sp \mu$
transforming as scalars, the velocities $q\sp \mu$ as vectors and the Lagrange
multipliers $\lambda\sp \mu$ as pseudo-scalars. In fact, due to this local
symmetry and the presence of Lagrange multipliers, (2.1) defines a system
whose Hamiltonian formulation includes constraints on
phase space$\sp {[22]}$. For generic rigid particles, one ends up with the
following description$\sp {[6,20]}$. Phase space degrees of freedom are the
coordinates $x\sp \mu$ and $q\sp \mu$ and their respective conjugate momenta
$$P_\mu={\pd L\over\pd \dot x\sp \mu}\ ,\quad
Q_\mu={\pd L\over\pd \dot q\sp \mu}\ \ .\eqno(2.6)$$
Here, $L\ts$ is the Lagrangian defining the action $\ts S=\int d\tau L\ts$
in (2.1). The associated symplectic structure is given by the canonical
Poisson brackets
$$\{x\sp \mu,P\sp \nu\}=\eta\sp {\mu\nu}\ \ ,\qquad
\{q\sp \mu,Q\sp \nu\}=\eta\sp {\mu\nu}\ \ .\eqno(2.7)$$
The sector related to the Lagrange multipliers decouples
when solving for the second-class constraints
$(\chi_1\sp \mu=\pi\sp \mu \equiv \pd L/\pd \dot\lambda_\mu)$ and
$(\chi_2\sp \mu=P\sp \mu - \mu c\lambda\sp \mu)$ using Dirac
brackets$\sp {[22]}$ (this is also true in the degenerate case).
Consequently, these degrees of freedom do not appear in the
Hamiltonian description. Actually, (2.1) being a spacetime Poincar\'e
invariant action, $P\sp \mu$ also defines the total energy-momentum of
the particle, while its total angular-momentum is given by
$$M\sp {\mu\nu}=L\sp {\mu\nu}+S\sp {\mu\nu}\ ,\eqno(2.8)$$
with the orbital and internal spin contributions, respectively
$$L\sp {\mu\nu}=P\sp \mu x\sp \nu -P\sp \nu x\sp \mu\ ,\quad
S\sp {\mu\nu}=Q\sp \mu q\sp \nu -Q\sp \nu q\sp \mu\ .\eqno(2.9)$$
With the brackets (2.7), $P\sp \mu$ and $M\sp {\mu\nu}$ of course obey the
Poincar\'e algebra. Both $Q\sp \mu$ and $S\sp {\mu\nu}$ are a measure of
the world-line extrinsic curvature, since (2.1) implies
$$Q\sp \mu=\ -2\mu c\kappa\ {F'(\kappa\sp 2 K\sp 2)\over\sqrt{-q\sp 2}}
\ \kappa K\sp \mu\ ,\quad S\sp {\mu\nu}=\ 2\mu c\kappa\ F'(\kappa\sp 2 K\sp 2)
\ \kappa(n\sp \mu K\sp \nu - n\sp \nu K\sp \mu)\ \ .\eqno(2.10)$$
These extrinsic curvature contributions suggest the possibility
that rigid particles could actually provide a unifying scheme for quantum
particles of different spins. In fact, assuming for a moment that this were
indeed the case, only integer spins would be obtained. Indeed, the
Heisenberg algebra of commutators
$\bigl[q\sp \mu,Q\sp \nu\bigr]=i\hbar\eta\sp {\mu\nu}$ associated to (2.7)
only supports single valued wave-function representations thereby
always leading to integer spin representations. However, the quantum
anomaly described later on shows that even such a possibility is
unfortunately not tenable.

The above phase space $\{ x\sp \mu, P_\mu; q\sp \mu, Q_\mu\}$ is subject
to the two first-class constraints
$$\phi_1=\ qQ\ ,\qquad \phi_2=\ qP+\mu c \sqrt{-q\sp 2}\Phi(q\sp 2 Q\sp 2)
\ ,\eqno(2.11)$$
with the Poisson bracket
$$\{\phi_1,\phi_2\}=\ -\phi_2\ .\eqno(2.12)$$
Here, $\Phi(q\sp 2 Q\sp 2)$ is given by
$$\Phi(q\sp 2 Q\sp 2)=\ F(x_0)-2 x_0 F'(x_0)\ ,\eqno(2.13)$$
with $x_0$ a solution to the equation
$$x_0 F'\sp 2(x_0)=\ {-q\sp 2 Q\sp 2\over (2\mu c\kappa)\sp 2} > 0
\ \ .\eqno(2.14)$$
Therefore, in order to define the Hamiltonian description
of generic rigid particles, the function $F(x)$ must also be
such that given any $y>0$ there always exists a unique $x>0$ for which
$x F'\sp 2(x)=y$. This condition puts some restriction on the class of
acceptable functions $F(x)$ in (2.1), which is assumed to be met in
our analysis. However, one may also take the point of view that the
Hamiltonian formulation is not necessarily directly related to the
Lagrangian one in (2.1), in which case only $\Phi(q\sp 2 Q\sp 2)$ needs
to be given and may be assumed to be any arbitrary {\it non constant}
function (\ts $\Phi(q\sp 2 Q\sp 2)$ constant corresponds to the degenerate
rigid particle). Finally, the total Hamiltonian for the system is simply
$$H=\ \lambda_1\phi_1 + \lambda_2\phi_2\ ,\eqno(2.15)$$
with $\lambda_1$ and $\lambda_2$ being Lagrange multipliers for the two
first-class constraints.

As is always$\sp {[22]}$ the case with first-class constraints, $\phi_1$
and $\phi_2$ are generators of local (Hamiltonian) gauge symmetries.
The transformations associated to $\phi_1$ -- a constraint equivalent
to the property $n K=0$ in (2.4) -- are simply$\sp {[20]}$ local rescalings
of the variables $q\sp \mu, Q\sp \mu$ and $\lambda_2$ and a local shift in
$\lambda_1$. On the other hand, local world-line reparametrisations
preserving the world-line orientation are generated$\sp {[20]}$ by the
constraint $\phi_2$ (to which a specific contribution from the constraint
$\phi_1$ must also added). In fact, the combination
$(\lambda_3=\lambda_1 + \dot\lambda_2/\lambda_2)$ defines$\sp {[20]}$ an
intrinsic world-line metric through $(\lambda_3=\dot e/e)$ with $e(\tau)$
being the world-line einbein. Using the freedom offered by these gauge
symmetries, it is always possible to choose for the Lagrange multipliers
$$\lambda_1=0\ ,\quad\lambda_2=1\ .\eqno(2.16)$$
Indeed, modular space$\sp {[22]}$ for this system reduces$\sp {[20]}$
to a single point. The configuration (2.16) is gauge equivalent to
this single point and actually defines$\sp {[20]}$ an
admissible$\sp {[22]}$, {\it i.e.} a complete and global gauge fixing
of the system.
\vskip 20pt
\leftline{\bf 3. The Unrestricted Phase Space Map}
\vskip 15pt
Naively, canonical quantisation of rigid particles would
proceed from their Hamiltonian formulation above. Heisenberg commutation
relations for the fundamental degrees of freedom would simply follow
from the Poisson brackets (2.7) and in the associated representation
space of quantum states  -- necessarily  equivalent to a
wave-function representation --, physical states would be identified
as being those states annihilated by two quantum operators in direct
correspondence with the first-class constraints $\phi_1$ and $\phi_2$
for some consistent choice of normal ordering of composite operators.
However, this approach -- the one so far always adopted for rigid
particles$\sp {[5,6,18,19]}$ -- overlooks one important feature concerning
the degrees of freedom $q\sp \mu$, namely the fact that this sector of
phase space is restricted by the requirement $(-q\sp 2 > 0)$ or in other
words that $q\sp \mu$ must lie {\it inside} the light-cone. Hence, in the
same way that the canonical quantisation of the nonrelativistic particle
moving freely on the positive real axis needs some specification$\sp {[23]}$,
we must first find a (canonical) transformation for the restricted
degrees of freedom $q\sp \mu$ and $Q\sp \mu$ such that the new set
of variables is unrestricted and preferably, is also equipped with
a canonical symplectic structure. Then, in terms of the transformed
degrees of freedom, the above quantisation program may be applied.

Such a set of transformed degrees of freedom indeed exists for
rigid particles. It is defined by the following relations
$$\eqalignno{y\sp 0=\ \eta\sqrt{-q\sp 2}\ ,\qquad
&R\sp 0=\ \eta\ {\bigl[-q\sp 0 Q\sp 0 +
\vec q. \vec Q \ \bigr]\over\sqrt{-q\sp 2}}\ ,&(3.1a)\cr
y\sp i=\ \eta\ {q\sp i\over\sqrt{-q\sp 2}}\ ,\qquad
&R\sp i=\ \eta\sqrt{-q\sp 2}\bigl[Q\sp i-{q\sp i\over q\sp 0}\ Q\sp 0\bigr]
\ ,&(3.1b)\cr}$$
where $\eta$ is the sign of $q\sp 0$ and $(i=1,2,\cdots,D-1)$ are space
indices. The inverse relations are
$$\eqalignno{q\sp 0=\ y\sp 0\sqrt{1+{\vec y}\ts\sp 2}\ ,\qquad
&Q\sp 0=\sqrt{1+{\vec y}\ts\sp 2}
\bigl[\ -R\sp 0+{{\vec y}.{\vec R}\over y\sp 0}\ \bigr]\ ,&(3.2a)\cr
q\sp i=\ y\sp 0\ts y\sp i\ ,\qquad
&Q\sp i=\ {R\sp i\over y\sp 0} + y\sp i
\bigl[\ -R\sp 0+{{\vec y}.{\vec R}\over y\sp 0}\ \bigr]\ . &(3.2b)\cr}$$
In geometrical terms, $y\sp 0$ measures the invariant length of the vector
$q\sp \mu$ with a sign related to whether $q\sp \mu$ lies in the forward
or in the backward light-cone, while the remaining variables $y\sp i$ are
in fact the parameters of the Lorentz boost in the direction $\vec q\ts$
mapping the vector $q\sp \mu=(q\sp 0,\vec q\ts)$ into
the vector $(y\sp 0,\vec 0\ts)$. The remaining variables $R\sp 0$ and
$R\sp i$ are then obtained as degrees of freedom conjugate to $y\sp 0$ and
$y\sp i$ respectively. Namely, the Poisson brackets (2.7) and the following
canonical brackets
$$\{y\sp 0,R\sp 0\}=1\ ,\quad \{y\sp i,R\sp j\}=\delta\sp {ij}\ ,\eqno(3.3)$$
are mapped into one another under the transformations (3.1) and (3.2).

Clearly, the canonically conjugate degrees of freedom
$(y\sp 0,R\sp 0;y\sp i,R\sp i)$ are no longer restricted as are the original
ones $(q\sp \mu,Q_\mu)$, thereby achieving the required properties. However,
the price to pay is a loss of manifest Lorentz covariance. Spacetime
translations generated by $P\sp \mu$ and space rotations generated by
$M\sp {ij}=L\sp {ij}+S\sp {ij}$
are still manifest symmetries in the transformed representation of the
system, but this is no longer the case for Lorentz boost generators
$M\sp {0i}=L\sp {0i}+S\sp {0i}$. Indeed, while the expressions (2.9) for
$L\sp {\mu\nu}$ are not affected by the redefinitions (3.2), those for the
spin tensor become
$$S\sp {0i}=\ -R\sp i\ts\sqrt{1+{\vec y}\ts\sp 2}\ ,\quad
S\sp {ij}=\ R\sp i y\sp j - R\sp j y\sp i\ .\eqno(3.4)$$
Nevertheless, it is a straightforward calculation to check that with the
brackets (3.3), the full Poincar\'e algebra is still obtained for the
generators $P\sp \mu$ and $M\sp {\mu\nu}$ expressed in terms of the
transformed variables (3.1), thereby establishing the consistency of this
alternative Hamiltonian description of rigid particles (the
redefinitions (3.1) are of
course also applicable in the degenerate case). In the generic case, the
first-class constraints (2.11) and associated Hamiltonian (2.15) are
then given by
$$\phi_1=\ y\sp 0\ts R\sp 0\ ,\eqno(3.5)$$
and
$$\phi_2=\ y\sp 0\bigl[{\vec y}.{\vec P} - P\sp 0\sqrt{1+
{\vec y}\ts\sp 2}\bigr]
+\eta\ts\mu c\ y\sp 0 \Phi\bigl(\ts(y\sp 0 R\sp 0)\sp 2 -
({\vec y}.{\vec R}) \sp 2 -{\vec R}\sp 2\bigr)\ ,\eqno(3.6)$$
with
$$q\sp 2 Q\sp 2=\ (y\sp 0 R\sp 0)\sp 2 -
({\vec y}.{\vec R})\sp 2 -{\vec R}\sp 2\ .
\eqno(3.7)$$
 From these expressions and the brackets (3.3), the gauge algebra (2.12) is
obviously also recovered.
\vskip 20pt
\leftline{\bf 4. The Mixed Lorentz-Gravitational Anomaly}
\vskip 15pt
The quantisation of rigid particles is thus specified by the Heisenberg
commutation relations
$$\bigl[x\sp \mu, P\sp \nu\bigr]=\ i\hbar\ \eta\sp {\mu\nu}\ ,\quad
\bigl[y\sp 0, R\sp 0\bigr]=\ i\hbar\ ,\quad \bigl[y\sp i,R\sp j\bigr]=
\ i\hbar\ \delta\sp {ij}\ ,\eqno(4.1)$$
and an abstract representation space of this algebra equipped
with an inner product for which these operators are all hermitian and
self-adjoint. Representations of this algebra are unitarily
equivalent to wave-function ones either in position or in momentum space
for each pair of conjugate degrees of freedom. This determines the
space of quantum states for such systems, each of these states being
therefore of positive norm.

Turning to the ordering problem, let us first
consider the situation for the Poincar\'e generators. Clearly, $P\sp \mu$ does
not require an ordering prescription. For $L\sp {\mu\nu}$ and
$S\sp {\mu\nu}$ we choose
$$L\sp {\mu\nu}=\ P\sp \mu x\sp \nu - P\sp \nu x\sp \mu\ ,\eqno(4.2)$$
and
$$S\sp {0i}=\ -{1\over 2}\bigl[\ R\sp i\sqrt{1+{\vec y}\ts\sp 2} +
\sqrt{1+{\vec y}\ts\sp 2}\ R\sp i\ \bigr]\ ,\quad S\sp {ij}=
\ R\sp i y\sp j - R\sp j y\sp i\ ,\eqno(4.3)$$
in order that these operators be hermitian and self-adjoint. Obviously,
$L\sp {\mu\nu}$ and $P\sp \mu$ generate the Poincar\'e algebra. On the
other hand, while it is clear that $S\sp {ij}$ generates the algebra of
rotations in space, it is not difficult to check that with the choice
of ordering in (4.3) the operators $S\sp {ij}$ and $S\sp {0i}$ in fact
obey the whole Lorentz algebra. Thus, the total angular-momentum
$(M\sp {\mu\nu}=L\sp {\mu\nu}+S\sp {\mu\nu})$ and energy-momentum $P\sp \mu$
operators generate the whole Poincar\'e algebra, thereby establishing
that this algebra is anomaly free and that in spite of
the loss of manifest spacetime covariance, quantum
states of rigid particles indeed span a linear representation space for
spacetime translations and Lorentz transformations. As already pointed out
above, this space can only support integer spin representations of the
Lorentz group.

Let us now turn to the ordering problem for the first-class constraints
$\phi_1$ and $\phi_2$, a necessary prerequisite in order to define
{\it physical} states of quantum rigid particles. Again, in order to
have hermitian and self-adjoint operators, we must choose for the quantum
constraints
$$\phi_1=\ {1\over 2}\bigl[\ y\sp 0 R\sp 0 + R\sp 0 y\sp 0\ \bigr]
\ ,\eqno(4.4a)$$
and
$$\eqalignno{
\phi_2=\ y\sp 0\bigl[{\vec y}.{\vec P} - P\sp 0&\sqrt{1+
{\vec y}\ts\sp 2}\bigr]+
{1\over 2}\ \eta\ts\mu c\ \bigl[y\sp 0\Phi\bigl(
:(y\sp 0 R\sp 0)\sp 2: - :({\vec y}.{\vec R})\sp 2:
- {\vec R}\sp 2\bigr) +\cr
&+\Phi\bigl(:(y\sp 0 R\sp 0)\sp 2: - :({\vec y}.{\vec R})\sp 2:
- {\vec R}\sp 2\bigr)\ y\sp 0\ \bigr]\ ,&(4.4b)\cr}$$
where $:(y\sp 0 R\sp 0)\sp 2:$ and $:({\vec y}.{\vec R})\sp 2:$ stand for
normal ordered expressions of the corresponding operators to be specified
presently. By considering all possible orderings for the products in these
operators, one concludes that the most general choices possible all reduce to
expressions of the following form
$$\eqalignno{:(y\sp 0 R\sp 0)\sp 2:&=\ R\sp 0 y\sp 0 y\sp 0 R\sp 0
+ i\hbar\ A_1\ y\sp 0 R\sp 0 + \hbar\sp 2 A_2\ ,&(4.5a)\cr
:({\vec y}.{\vec R})\sp 2:&=\ R\sp i y\sp i y\sp j R\sp j
+ i\hbar\ B_1\ y\sp i R\sp i + \hbar\sp 2 B_2\ ,&(4.5b)\cr}$$
where $A_1, A_2, B_1$ and $B_2$ are undetermined free complex coefficients.
Requiring that these operators be also hermitian and self-adjoint only
leads to the restrictions
$$\eqalignno{A_1\sp *=\ - A_1\ ,&\quad A_2\sp *=\ A_2 - A_1\ ,&(4.6a)\cr
B_1\sp *=\ - B_1\ ,&\quad B_2\sp *=\ B_2 - (D-1) B_1\ .&(4.6b)\cr}$$

With these definitions, it is now possible to determine the commutation
relations for the quantum gauge algebra. One easily finds
$$\bigl[\ \phi_1,\ \phi_2\ \bigr]=\ -i\hbar\ \phi_2\ .\eqno(4.7)$$
Comparison with the classical bracket (2.12) shows that the gauge algebra is
indeed anomaly free. Therefore, both local world-line reparametrisations
and the local rescalings generated by $\phi_1$ are symmetries of quantised
generic rigid particles. From that point of view, it is thus
meaningful to define their quantum {\it physical} states $|\psi>$
as being the solutions to the conditions
$$\phi_1 |\psi>\ =\ 0\ ,\qquad \phi_2 |\psi>\ =\ 0\ ,\eqno(4.8)$$
thereby ensuring invariance of these states under all local gauge
symmetries including world-line reparametrisations. However, this
definition must also be consistent with the other symmetries of the system.
Namely, the generators of gauge symmetries must commute with those of
spacetime Poincar\'e transformations. Otherwise, {\it physical} states
solving (4.8) cannot define linear representations of the Poincar\'e group.
In other words, a state physical in a given reference frame would no
longer be physical in some other frame! Nor would it be possible to
define consistently the mass or spin of {\it physical} states!

Clearly, this type of problem does not arise for the gauge
generator $\phi_1$ since
$$\eqalignno{\bigl[\ L\sp {\mu\nu},\ \phi_1\bigr]=0\ ,\quad
\bigl[\ S\sp {\mu\nu},&\ \phi_1\bigr]=0\ ,\quad
\bigl[\ M\sp {\mu\nu},\ \phi_1\bigr]=0\ ,\cr
\bigl[\ P\sp \mu ,&\ \phi_1 \bigr]=0\ .&(4.9)\cr}$$
Moreover, we also have for the generator of world-line reparametrisations
$$\bigl[\ P\sp \mu ,\ \phi_2 \bigr]=0\ .\eqno(4.10)$$
Therefore, at least the energy-momentum hence also the mass of quantum
physical states are well defined observables for generic rigid particles.
To analyse the situation for the remaining commutators
$\bigl[M\sp {\mu\nu},\phi_2 \bigr]$, it is useful to decompose
$\phi_2$ in (4.4b) as $\phi_2=\chi_1+\chi_2$ with
$$\chi_1=\ y\sp 0\bigl[{\vec y}.{\vec P} - P\sp 0
\sqrt{1+{\vec y}\ts\sp 2}\bigr]\ .\eqno(4.11)$$
A simple calculation then finds that
$$\eqalignno{\bigl[\ L\sp {0i},\ \chi_1\bigr]
=&\ i\hbar (P\sp 0 y\sp 0 y\sp i - P\sp i y\sp 0\sqrt{1+{\vec y}\ts\sp 2})
=\ - \bigl[S\sp {0i},\ \chi_1\bigr]\ ,&(4.12a)\cr
\bigl[\ L\sp {ij},\ \chi_1\bigr]
=&\ i\hbar (P\sp i y\sp 0 y\sp j - P\sp j y\sp 0 y\sp i)
=\ -\bigl[\ S\sp {ij},\ \chi_1\bigr]\ ,&(4.12b)\cr}$$
leading to
$$\bigl[\ M\sp {\mu\nu},\ \chi_1\bigr]=\ 0\ ,\eqno(4.13)$$
and
$$\bigl[\ M\sp {\mu\nu},\ \phi_2\bigr]=
\bigl[\ M\sp {\mu\nu},\ \chi_2\bigr]\ .\eqno(4.14)$$
Moreover, since $L\sp {\mu\nu}$ clearly also commutes with $\chi_2$,
only the commutators of $S\sp {\mu\nu}$ with $\chi_2$ are left to be computed.
In fact, since both $y\sp 0$ and $R\sp 0$ commute with $S\sp {\mu\nu}$,
the crucial commutators to be determined are those of $S\sp {\mu\nu}$ with
$\big(:(y\sp 0 R\sp 0)\sp 2: - :({\vec y}.{\vec R})\sp 2:
- {\vec R}\sp 2 \bigr)$.
Using the normal ordered expressions (4.5), a direct calculation shows that
$$\bigl[\ S\sp {ij},\ :(y\sp 0 R\sp 0)\sp 2: - :({\vec y}.{\vec R})\sp 2:
- {\vec R}\sp 2 \bigr]=\ 0\ ,\eqno(4.15)$$
so that finally
$$\bigl[\ M\sp {ij},\ \phi_2 \bigr]=0\ .\eqno(4.16)$$
This result is indeed to be expected owing to the manifest rotation
covariance of the quantisation procedure. On the other hand, the commutator
with Lorentz boost generators gives
$$\eqalignno{
\bigl[\ S\sp {0i},\ :(y\sp 0 R\sp 0)\sp 2: &\ - :({\vec y}.{\vec R})\sp 2:
- {\vec R}\sp 2 \bigr]=
\ {1\over 2}\ i\hbar\sp 3\ {y\sp i\over (1+{\vec y}\ts\sp 2)\sp {3/2}}\ +\cr
+\ {1\over 2}\ \hbar\sp 2\ B_1\ &\bigl[\ R\sp i{1\over\sqrt{1+
{\vec y}\ts\sp 2}}
+{1\over\sqrt{1+{\vec y}\ts\sp 2}} R\sp i\ \bigr]\ .&(4.17)\cr}$$
Hence, we certainly have for any choice of $F(\kappa\sp 2 K\sp 2)$ in the
generic case
$$\bigl[\ M\sp {0i},\phi_2\bigr] \neq 0\ .\eqno(4.18)$$
This is the mixed Lorentz-gravitational anomaly of the title. Generally,
this anomaly is of order $\hbar\sp 2$ unless an ordering for $\phi_2$
corresponding to $B_1=0$ in (4.5b) happens to be chosen, in which case
the anomaly is of order $\hbar\sp 3$. Therefore, given {\it any\ts}
ordering for the generator of local world-line reparametrisations,
physical states (4.8) do {\it not\ts} transform covariantly under Lorentz
boosts! The subspace of physical states (4.8) is not closed under the
action of Lorentz generators, even though these generators act covariantly
on the {\it entire} space of states.
\vskip 20pt
\leftline{\bf 5. Conclusions}
\vskip 15 pt
This letter has described how by properly accounting for the restriction of
causal propagation inside the light-cone, rigid particles cannot be quantised
in a way which is compatible with the requirements of local gauge invariance
under world-line reparametrisations and of spacetime Poincar\'e covariance
both at the same time. The origin of the problem lies with a mixed
Lorentz-gravitational anomaly in the commutator of Lorentz boosts with
the generator of world-line reparametrisations. Consequently, even though
the Poincar\'e and local gauge algebras are both anomaly free on the
{\it complete} space of states, Lorentz boosts map {\it outside} of the
{\it subspace} of {\it physical} states defined to be all states invariant
under gauge transformations -- which includes world-line reparametrisations.
In fact, the only Poincar\'e invariant quantum observable which is well
defined for physical states is their mass; the concept of spin has no
meaning for these states. We must therefore conclude that rigid particles
are {\it not} consistent quantum models for particle physics.
To be precise, the present analysis applies for any possible dependence
on the world-line extrinsic curvature with only one exception, corresponding
to a degenerate situation for which classical solutions are all of
constant extrinsic curvature. This degenerate case requires a separate
treatment still to be completed. Nevertheless, the same type of anomaly
as the one above would presumably be obtained in that case as well.
Most probably, the same conclusion would extend further to theories which
also include a dependence on the extrinsic torsion and other such
invariants of higher order still.

One may also argue that the quantum anomaly for rigid particles is the
strongest indication yet as to the probable inconsistency of quantised
rigid strings. It is widely believed$\sp {[24]}$ that the higher
derivative couplings of rigid string theories lead to quantum physical
states either of negative norm or of energy unbounded below. In fact, a
semi-classical analysis of Polyakov's rigid strings has revealed$\sp {[25]}$
instabilities of the latter type. However, rigid strings
possess collapsed configurations corresponding to rigid particles.
Since quantised rigid particles are not consistent, quantised rigid
strings cannot be consistent either. Note that quantum inconsistency
of rigid particles is not related either to negative-norm
physical states nor to energy unbounded below but actually follows from
a quantum anomaly. Strictly speaking, if this type of reasoning is
justified, the conclusion applies so far only to those rigid strings
whose collapsed configurations are not degenerate rigid particles.
Specifically, consider the dimensional reduction$\sp {[26,8]}$ of a
$(D+1)$-dimensional rigid string whose action depends on the
world-sheet extrinsic curvature through some function $G(x)$,
$$S[\phi\sp M]=\ -{\mu c\over\kappa}\int d\tau\ d\sigma\ \sqrt{-g}
\ G\bigl(\kappa\sp 2 \triangle\phi\sp M \triangle\phi_M\bigr)\ .\eqno(5.1)$$
Here, $\phi\sp M (M=0,1,\cdots,D)$ are the string coordinates,
$g_{\alpha\beta}=\eta_{MN}\pd_\alpha\phi\sp M\pd_\beta\phi\sp N$ is the
induced world-sheet metric ($\eta_{MN}$ is the Minkowski metric in $(D+1)$
dimensions), $\triangle$ is the Laplacian
$$\triangle\ =\ {1\over\sqrt{-g}}\ \pd_\alpha\ \sqrt{-g}
\ g\sp {\alpha\beta} \pd_\beta\ \ ,\eqno(5.2)$$
and as usual $\xi\sp {\alpha=0}=\tau$ and $\xi\sp {\alpha=1}=\sigma$
with $\alpha,\beta=0,1$. When identifying$\sp {[26,8]}$ one of the space
coordinates $\phi\sp M$ with $\sigma$ and assuming that the remaining
string coordinates $x\sp \mu$ are independent of $\sigma$, (5.2)
reduces to (2.1) with
$$F(x)\ =\ G(x)\ \int d\sigma \eqno(5.3)$$
(the integral is over the finite range of $\sigma$). Thus in particular,
Polyakov's rigid strings$\sp {[1,8]}$ correspond to the choice
$$F(x)=\alpha_0\ts x + \beta_0\ .\eqno(5.4)$$
Since this function does not define the degenerate case (2.5),
we must conclude from the analysis of this paper that Polyakov's rigid
strings cannot be consistent fundamental quantum theories. Of course,
this does not necessarily exclude their possible relevance as effective
theories for some {\it semi-classical} approximation to other more
fundamental theories.
\vskip 20 pt
\leftline{\bf Acknowledgement}
\vskip 15pt
This work was supported through a Senior Research Assitant
position funded by the S.E.R.C.
\vskip 20pt
\leftline{\bf REFERENCES}
\vskip 15pt
\frenchspacing
\item{[1]} A. M. Polyakov, Nucl. Phys. {\bf B268} (1986) 406.
\item{[2]} H. Kleinert, Phys. Lett. {\bf B174} (1986) 335.
\item{[3]} R. D. Pisarski, Phys. Rev. {\bf D34} (1986) 670.
\item{[4]} C. Battle, J. Gomis and N. Roman-Roy,
J. Phys. {\bf A21} (1988) 2693.
\item{[5]} V. V. Nesterenko, J. Phys. {\bf A22} (1989) 1673;
Theor. Math. Phys. {\bf 86} (1991) 168; Mod. Phys. Lett. {\bf A6} (1991) 719.
\item{[6]} M. S. Plyushchay, Mod. Phys. Lett. {\bf A3} (1988) 1299;
{\it ibid} {\bf A4} (1989) 837;
{\it ibid} {\bf A4} (1989) 2747; Int. J. Mod. Phys. {\bf A4} (1989) 3851;
Phys. Lett. {\bf B235} (1990) 47; {\it ibid} {\bf B236} (1990) 291;
{\it ibid} {\bf B243} (1990) 383; {\it ibid} {\bf B248} (1990) 107;
{\it ibid} {\bf B248} (1990) 299; {\it ibid} {\bf B253} (1991) 50;
{\it ibid} {\bf B262} (1991) 71; {\it ibid} {\bf B280} (1992) 232;
Nucl. Phys. {\bf B362} (1991) 54.
\item{[7]} A. Dhar, Phys. Lett. {\bf B214} (1988) 75.
\item{[8]} J. Grundberg, J. Isberg, U. Lindstr\"om and H. Nordstr\"om,
Phys. Lett. {\bf B231} (1989) 61.
\item{[9]} J. P. Gauntlett, K. Itoh and P. K. Townsend,
Phys. Lett. {\bf B238} (1990) 65.
\item{[10]} M. Pav\v si\v c, Phys. Lett. {\bf B205} (1988) 231;
{\it ibid} {\bf B221} (1989) 264.
\item{[11]} M. Huq, P. I. Obiakor and S. Singh,
Int. J. Mod. Phys. {\bf A5} (1990) 4301.
\item{[12]} H. Arodz, A. Sitarz and P. Wegrzyn,
Acta Phys. Polonica {\bf B20} (1989) 921.
\item{[13]} T. Dereli, D. H. Harley, M. \"Onder and R. W. Tucker,
Phys. Lett. {\bf B252} (1990) 601.
\item{[14]} D. Zoller, Phys. Rev. Lett. {\bf 65} (1990) 2236.
\item{[15]} G. Fiorentini, M. Gasperini and G. Scapetta,
Mod. Phys. Lett. {\bf A6} (1991) 2033.
\item{[16]} A. M. Polyakov, Mod. Phys. Lett. {\bf A3} (1988) 325.
\item{[17]} S. Iso, C. Itoi and H. Mukaida,
Phys. Lett. {\bf B236} (1990) 287; Nucl. Phys. {\bf B346} (1990) 293.
\item{[18]} Yu. A. Kuznetsov and M. S. Plyushchay, {\sl ``The Model of
the Relativistic Particle with Curvature and Torsion''},
Protvino preprint IHEP 91-162 (October 1991), and references therein.
\item{[19]} J. Isberg, U. Lindstr\"om and H. Nordstr\"om,
Mod. Phys. Lett. {\bf A5} (1990) 2491.
\item{[20]} J. Govaerts, in preparation.
\item{[21]} L. Alvarez-Gaum\'e and E. Witten,
Nucl. Phys. {\bf B234} (1983) 269.
\item{[22]} For a recent review, see \hfil\break
J. Govaerts, {\it Hamiltonian Quantisation and Constrained Dynamics},
Lecture Notes in Mathematical and Theoretical Physics {\bf 4},
(Leuven University Press, Leuven, 1991).
\item{[23]}C. J. Isham, in {\it Relativity, Groups and Topology II},
Les Houches 1983, eds. B. S. DeWitt and R. Stora (North Holland,
Amsterdam, 1984), p. 1162.
\item{[24]} See for example \hfil\break
J. Polchinski and Z. Yang, {\sl ``High Temperature Partition
Function of the Rigid String''}, Texas/Rochester preprint UTTG-08-92,
UR-1254, ER-40685-706.
\item{[25]} E. Braaten and C. K. Zachos, Phys. Rev. {\bf D34} (1987) 1512.
\item{[26]} M. J. Duff, P. S. Howe, T. Inami and K. S. Stelle,
Phys. Lett. {\bf B191} (1987) 70.
\end